\documentclass[a4paper,USenglish,cleveref,pdfa,autoref]{lipics-v2021}

\pdfoutput=1 %
\hideLIPIcs  %

\title{Certifying Higher-Order Polynomial Interpretations}

\author{Niels van der Weide}
{Institute for Computing and Information Sciences, Radboud University, Nijmegen, The Netherlands \and \url{https://nmvdw.github.io}}
{nweide@cs.ru.nl}
{https://orcid.org/0000-0003-1146-4161}
{This research was supported by the NWO project
``The Power of Equality'' OCENW.M20.380,
which is financed by the Dutch Research Council (NWO).}

\author{Deivid Vale}
{Institute for Computing and Information Sciences, Radboud University, Nijmegen, The Netherlands \and \url{https://deividrvale.github.io}}
{D.Vale@cs.ru.nl}
{https://orcid.org/0000-0003-1350-3478}
{Author supported by NWO Top project
``Implicit Complexity through Higher-Order Rewriting'', NWO 612.001.803/7571.}

\author{Cynthia Kop}
{Institute for Computing and Information Sciences, Radboud University, Nijmegen, The Netherlands \and \url{https://www.cs.ru.nl/~cynthiakop/index_en.html}}
{C.Kop@cs.ru.nl}
{https://orcid.org/0000-0002-6337-2544}
{Author supported by NWO Top project
``Implicit Complexity through Higher-Order Rewriting'', NWO 612.001.803/7571
and the NWO VIDI project ``Constrained Higher-Order Rewriting and Program Equivalence'', NWO VI.Vidi.193.075.}

\authorrunning{N. van der Weide, D. Vale, and C. Kop}

\Copyright{Niels van der Weide, Deivid Vale, and Cynthia Kop}



\category{} %
\relatedversiondetails[cite=vanderweide_et_al:LIPIcs.ITP.2023.30]
{Published Conference Version}
{https://drops.dagstuhl.de/opus/volltexte/2023/18405} %

\supplementdetails[
  linktext={nmvdw/Nijn},
  subcategory={GitHub repository},
  cite={weide:vale:22}
]
{Coq Formalization}
{https://github.com/nmvdw/Nijn}

\supplementdetails[
    linktext={deividrvale/nijn-coq-script-generation},
    subcategory={GitHub repository},
    cite={vale:weide:23}
]
{Proof script generator}
{https://github.com/deividrvale/nijn-coq-script-generation}

\acknowledgements{The authors thank Dan Frumin for his help with understanding and using \lstinline{Ltac}.}

\nolinenumbers %

\EventEditors{John Q. Open and Joan R. Access}
\EventNoEds{2}
\EventLongTitle{42nd Conference on Very Important Topics (CVIT 2016)}
\EventShortTitle{CVIT 2016}
\EventAcronym{CVIT}
\EventYear{2016}
\EventDate{December 24--27, 2016}
\EventLocation{Little Whinging, United Kingdom}
\EventLogo{}
\SeriesVolume{42}
\ArticleNo{23}

\usepackage{amsmath,amssymb,amsfonts,amscd,amsbsy,mathtools}
\usepackage{bussproofs}
\usepackage{bm}

\usepackage{xcolor}
\usepackage{xcolor-material}

\newcommand{\consFont}[1]{\mathsf{#1}}
\newcommand{\defFont}[1]{\mathsf{#1}}

\newcommand{\metaFont}[1]{\mathtt{#1}}

\newcommand{\nil}{\consFont{nil}}

\newcommand{\cons}{\consFont{cons}}

\newcommand{\map}{\defFont{map}}

\newcommand{\atype}{A}
\newcommand{\btype}{B}

\newcommand{\avar}{x}

\newcommand{\aListVar}{xs}

\newcommand{\aFuncVar}{F}

\newcommand{\aterm}{s}

\newcommand{\rules}{\mathcal{R}}
\newcommand{\Nat}{\mathbb{N}}

\newcommand{\arrz}{\to}

\newcommand{\interpret}[1]{\llbracket #1\rrbracket}
\newcommand{\typeinterpret}[1]{\llparenthesis #1 \rrparenthesis}

\newcommand{\costGt}{>}
\newcommand{\costGe}{\geq}

\NewDocumentCommand{\typeVec}{m o}{
    \IfValueTF{#2}{
        {}^{#2}\bm{#1}
    }{
        \bm{#1}
    }
}

\newcommand{\app}{\,}

\newcommand{\hasType}{\mathbin{:}}

\NewDocumentCommand{\irc}{o}{
    \IfValueTF{#1}{
        \metaFont{irc}_{\rules_{#1}}
    }{
        \metaFont{irc}_{\rules}
    }
}

\newcommand{\arrtype}{\Rightarrow}
\newcommand{\arrfunc}{\longrightarrow}
\newcommand{\arrfuncwm}{\rightarrow_{\text{wm}}}

\newcommand{\defeq}{\coloneqq}

\newcommand{\pol}[1]{\consFont{Pol}(#1)}
\newcommand{\polTy}[2]{\consFont{Pol}^{#2}(#1)}

\newcommand{\minEl}[1]{\bot_{#1}}
\newcommand{\lvf}[1]{\consFont{lvf}_{#1}}

\usepackage{amsthm}
\usepackage{mathtools}
\usepackage{url}
\usepackage{xspace}
\usepackage[shortcuts]{extdash}
\usepackage{turnstile}
\usepackage{stmaryrd}
\usepackage{dashbox}
\usepackage{mathtools}
\usepackage{newunicodechar}
\usepackage{xifthen}

\usepackage[numbers,sort]{natbib}

\theoremstyle{definition}
\newtheorem{defi}{Definition}[section]
\newtheorem{exa}[defi]{Example}
\newtheorem{rem}[defi]{Remark}

\theoremstyle{plain}
\newtheorem{thrm}[defi]{Theorem}
\newtheorem{prop}[defi]{Proposition}

\newcommand{\bye}[1]{}

\newcommand\isafor{\textsf{Isa\kern-0.15exF\kern-0.15exo\kern-0.15exR}}
\newcommand\ceta{\textsf{C\kern-0.15exe\kern-0.45exT\kern-0.45exA}}
\newcommand{\COLORlib}{\textsf{CoLoR}}
\newcommand{\coccinelle}{\textsf{Cochinelle}}
\newcommand{\rainbow}{\textsf{Rainbow}}
\newcommand{\nijn}{\textsf{Nijn}}
\newcommand{\onijn}{\textsf{ONijn}}

\newcommand{\matchbox}{\texttt{MatchBox}}
\newcommand{\cime}{CiME3}
\newcommand{\TTT}{\textsf{T\kern-0.15em\raisebox{-0.55ex}T\kern-0.15emT\kern-0.15em\raisebox{-0.55ex}2}}
\newcommand{\aprove}{\textsf{AProVE}}
\newcommand{\NATT}{\textsf{NaTT}}
\newcommand{\wanda}{\textsf{Wanda}}
\newcommand{\MuTerm}{\textsc{Mu-Term}}

\input{coq.sty}
\newunicodechar{⟶}{\( \longrightarrow \)}
\newunicodechar{∙}{\( \bullet \)}
\newunicodechar{λ}{\( \lambda \)}
\newunicodechar{·}{\( \cdot \)}
\newunicodechar{∼}{\( \sim \)}
\newunicodechar{∘}{\( \circ \)}
\newunicodechar{⟦}{\( \llbracket \)}
\newunicodechar{⟧}{\( \rrbracket \)}
\newunicodechar{₁}{\(_1\)}
\newunicodechar{₂}{\(_2\)}

\newcommand{\coqdocbasebaseurl}{https://nmvdw.github.io/Nijn/html}

\newcommand{\coqdocbaseurl}{\coqdocbasebaseurl /Nijn.}
\newcommand{\urlhash}{\#}
\newcommand{\coqdocurl}[2]{\coqdocbaseurl #1.html\urlhash #2}
\newcommand{\nolinkcoqident}[1]{\nolinkurl{#1}} %
\makeatletter
\newcommand{\coqident}{\begingroup\@makeother\#\@coqident}
\newcommand{\@coqident}[3][]{%
	\ifthenelse{\isempty{#2}}%
	{\nolinkcoqident{#3}}%
	{\ifthenelse{\isempty{#1}}%
		{\href{\coqdocurl{#2}{#3}}{\nolinkcoqident{#3}}}%
		{\href{\coqdocurl{#2}{#3}}{\nolinkcoqident{#1}}}}%
	\endgroup}
\newcommand{\coqfile}[2]{%
	{\href{\coqdocbaseurl #1.#2.html}{\nolinkcoqident{#2.v}}}}
\makeatother

\lstdefinestyle{custombash}{
  language=bash,
  basicstyle=\ttfamily\color{black},
  keywordstyle=[1]\color{green!40!black}\bfseries,
  keywordstyle=[2]\color{blue}\bfseries,
  stringstyle=\color{purple!40!black}\bfseries,
  commentstyle=\color{gray}\itshape,
  moredelim=[s][\color{blue}]{\$}{\ },
  moredelim=[s][\color{blue}]{\{}{\}},
  moredelim=[s][\color{blue}]{\[}{\]},
  moredelim=[s][\color{blue}]{(}{)},
  moredelim=[s][\color{blue}]{<}{>},
  moredelim=[is][\bfseries\color{blue!40!black}]{\{}{\}},
  moredelim=[is][\bfseries\color{blue!40!black}]{\[}{\]},
  moredelim=[is][\bfseries\color{green!40!black}]{(}{)},
  moredelim=[is][\bfseries\color{purple!40!black}]{<}{>},
}

\begin{document}

\maketitle

\begin{abstract}
Higher-order rewriting is a framework in which one can write higher-order
programs and study their properties.
One such property is termination: the situation that for all inputs,
the program eventually halts its execution and produces an output.
Several tools have been developed to check whether higher-order rewriting systems
are terminating.
However, developing such tools is difficult and can be error-prone.
In this paper, we present a way of certifying termination proofs of higher-order
term rewriting systems.
We formalize a specific method that is used to prove termination,
namely the polynomial interpretation method.
In addition,
we give a program that processes proof traces containing a high-level description
of a termination proof into a formal Coq proof script that can be checked
by Coq.
We demonstrate the usability of this approach by certifying higher-order
polynomial interpretation proofs produced by Wanda,
a termination analysis tool for higher-order rewriting.
\end{abstract}

\section{Introduction}\label{sec:introduction}

Automatically proving termination is an important problem in term rewriting,
and numerous tools have been developed for this purpose,
such as \aprove{}~\cite{jurgen:et-al:17},
\NATT{}~\cite{yamada:kusakari:sakabe:14},
\matchbox{}~\cite{waldmann:04},
\MuTerm{}~\cite{DBLP:conf/cade/GutierrezL20a},
SOL~\cite{hamana:20},
\TTT~\cite{korp:sternagel:zankl:09} and
\wanda{}~\cite{kop:20}, which compete against each other in an annual termination
competition~\cite{termcomp}. Aside from basic (first-order) term rewriting,
this includes tools analyzing for instance string, conditional,
and higher-order rewriting.

Developing termination tools is a difficult and error-prone endeavor.
On the one hand,
the termination techniques that are implemented may contain errors.
This is particularly relevant in higher-order term rewriting,
where the proofs are often very intricate due to partial application,
type structure, beta-reduction, and techniques often not transferring perfectly
between different formalisms of higher-order rewriting.
Hence, it should come as
no surprise that errors have been found even in published papers on higher-order
rewriting.
On the other hand,
it is very easy for a tool developer to accidentally omit a
test whether some conditions to apply specific termination techniques
are satisfied,
or to incorrectly translate a method between higher-order formalisms.

To exacerbate this issue, termination proofs are usually complex and
technical in nature,
which makes it hard to assess the correctness of a prover's output by hand.
Not only do many benchmarks contain hundreds of rules,
modern termination tools make use of various proof methods that have been
developed for decades.
Indeed,
a single termination proof might, for instance, make use of a
combination of
dependency~pairs~\cite{fuhs:kop:19,kop:raamsdonk:12,arts:00},
recursive~path~orders~\cite{koprowski:09,blanqui:jouannaud:rubio:15},
rule~removal,
and multiple kinds of
interpretations~\cite{fuhs:kop:12,neurauter:middeldorp:11,yamada:21,kop:vale:21}.
This makes bugs very difficult to find.

Hence, there is a need to formally certify the output of termination provers,
ideally automatically.
There are two common engineering strategies to provide such certification.
In the first,
one builds the certifier as a library in a proof assistant along with
tools to read the prover's output and construct a formal proof,
which we call \textit{proof script}.
The proof script is then verified by a proof assistant.
Examples of this system design are the combinations
{\coccinelle{}/\cime{}}~\cite{contejean_et_al:11} and
{\COLORlib{}/\rainbow{}}~\cite{blanqui:koprowski:11}.
In the second,
the formalization includes certified algorithms for checking the correctness of the prover's output.
This allows for the whole certifier to be extracted, using code extraction,
and be used as a standalone program.
Hence,
the generation of proof scripts by a standalone tool is not needed in this approach,
but it comes with a higher formalization cost.
{\isafor{}/\ceta{}}~\cite{rene:ster:09} utilizes
this approach.

When it comes to higher-order rewriting, however, the options are limited.
Both \coccinelle{}~\cite{contejean_et_al:11} and {\isafor{}/\ceta{}}~\cite{rene:ster:09}
only consider first-order rewriting.
{\COLORlib{}/\rainbow{}}~\cite{blanqui:koprowski:11} does include a
formalization of an early definition of HORPO~\cite{koprowski:09}.
Since here
we use a different term formalism compared to that of~\cite{koprowski:09},
our results are not directly compatible.
See for instance~\cite{almeida:ayala:20,reina:alonso:hidalgo:01}
for more formalization results in rewriting.

In this paper,
we introduce a new combination {\nijn{}/\onijn{}} for the certification of
higher-order rewriting termination proofs.
We follow the first aforementioned system design:
\nijn{} is a Coq library providing a formalization of the underlying higher-order
rewriting theory and \onijn{} is a proof script generator
that given a minimal description of a termination proof
(which we call \textit{proof trace}), outputs a Coq proof script.
The proof script then utilizes results from \nijn{}
for checking the correctness of the traced proof.
The schematic below depicts
the basic steps to produce proof certificates
using {\nijn{}/\onijn{}}.

\begin{figure}[h!]
    \includegraphics[width=\textwidth]{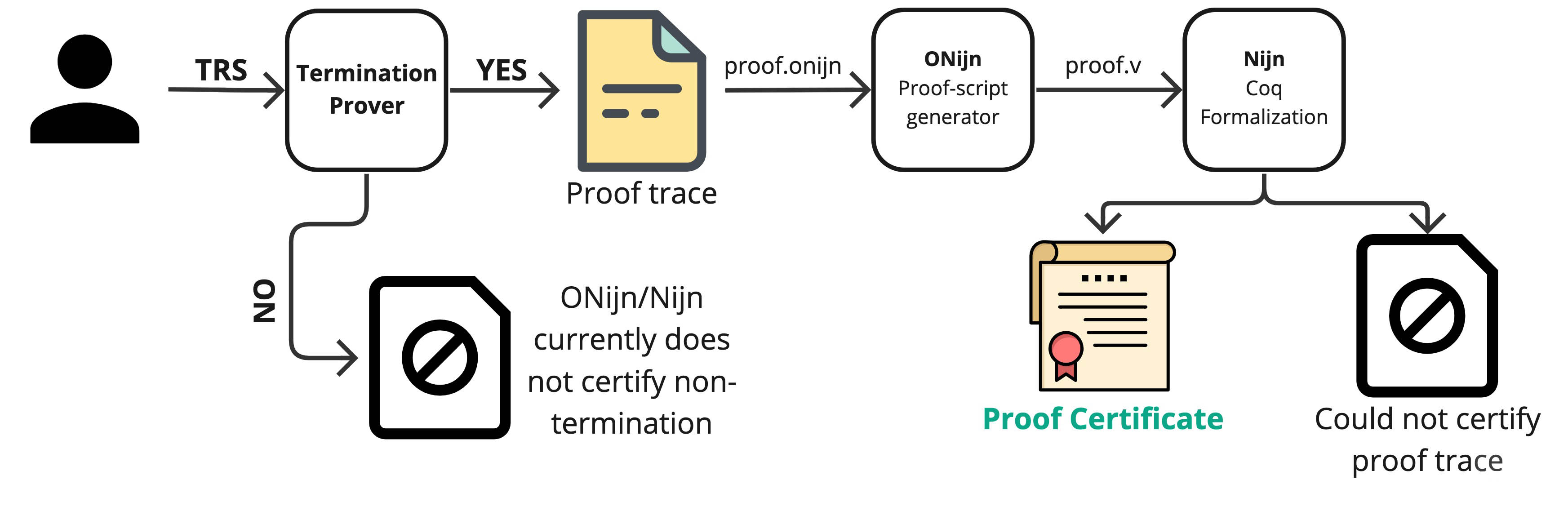}
    \caption{{\nijn{}/\onijn{}} schematics}\label{fig:onij-nijn-schm}
\end{figure}

While \nijn{} is the certified core part of our tool
since it is checked by Coq,
the proof script generation implemented in OCaml (\onijn{})
is not currently certified and must be trusted.
For this reason,
we deliberately keep \onijn{} as simple as possible
and no checking or computation is done by it.
The only task delegated to \onijn{} is that of parsing the proof trace given
by the termination prover to a Coq proof script.
Additionally,
checking the correctness of polynomial termination proofs in Coq
is an inherently incomplete task,
since it would require a method to solve inequalities over arbitrary
polynomials, which is undecidable in general.

\subparagraph*{Contributions.}
The main contribution of this paper can be summarized as follows:
\begin{itemize}
    \item we provide a formalization of higher-order
    algebraic functional systems (\cref{def:afs});
    \item a formal proof of the interpretation method using weakly monotonic
      algebras (\cref{thm:sn-interpretation});
    \item a formalization of the higher-order polynomial method (\cref{thm:poly-sn});
    \item a tactic that automatically solves the constraints that arise
    when using the higher-order polynomial method
    (\cref{sec:poly-notations});
    \item an OCaml program that transforms the output of a termination prover into a Coq script
    that represents the termination proof (\cref{sec:proof-scripts}).
\end{itemize}

\subparagraph*{Technical Overview.}
This paper orbits \nijn{}, a Coq library formalizing higher-order rewriting~\cite{weide:vale:22}.
The formalization is based on intensional dependent type theory
extended with two axioms:
\textit{function extensionality} and
\textit{uniqueness~of~identity~proofs}~\cite{DBLP:conf/lics/HofmannS94}.
Currently, the termination criterion formalized in the library is
\textit{the higher-order polynomial method}, introduced in~\cite{fuhs:kop:12}.
The tool \verb!coqwc! counts the following amount of lines of code:

\begin{verbatim}
spec    proof comments
5497     1985      272 total
\end{verbatim}

The higher-order interpretation method roughly works as follows.
First, types are interpreted as well-ordered structures (\cref{def:sem_Ty}),
compositionally.
For instance,
we interpret base types as natural numbers (with the usual ordering).
Then
we interpret a functional type \( \atype \arrtype \btype \) as
the set of weakly monotonic functions from \( \typeinterpret{\atype} \)
to \( \typeinterpret{\btype} \) where
\( \typeinterpret{\atype} \), \( \typeinterpret{\btype} \) denote the
interpretations of \( \atype, \btype \) respectively.
The second step is to map inhabitants of a type \( \atype \)
to elements of \( \typeinterpret{\atype} \),
which is expressed here by \cref{def:int-terms}.

This interpretation, called \textit{extended monotonic algebras} in~\cite{fuhs:kop:12}, alone does not suffice for termination.
To guarantee termination, we interpret both term~application (\cref{def:interpret-app}) and function symbols
as strongly monotonic functionals.
In addition, terms must be interpreted in such a way that the rules of the system are strictly oriented,
i.e.,
\( \interpret{\ell} \costGt \interpret{r} \),
for all rules \( \ell \arrz r \).
This means that whenever a rewriting is fired in a term,
the interpretation of that term strictly decreases.
As such, termination is guaranteed.
Here we use \textit{termination models} (\cref{def:term-model})
to collect these necessary conditions.

The main result establishing the correctness of this technique in the higher-order
case is expressed by \cref{thm:sn-interpretation}.
To the reader familiar with \textit{the interpretation method}
in first-order rewriting,
\cref{thm:poly-sn} would be no surprise.
It is essentially the combination of the Manna--Ness criterion
with higher-order polynomials and the additional technicalities
that are needed for the higher-order case.

\section{The Basics of Higher-Order Rewriting in Coq}\label{sec:preliminaries}
In this section,
we introduce the basic constructs needed to formalize
\textit{algebraic functional systems} (AFSs) like
types, contexts, variables, terms, and rewriting rules.
We end the section with an exposition on how to express termination
constructively in Coq.

\subsection{Terms and Rewrite Rules}

Let us start by defining \emph{simple types}.

\begin{defi}[\coqident{Syntax.Signature.Types}{ty}]
\label{def:stypes}
\textbf{Simple types} over a type \lstinline{B} are defined as follows:
\begin{lstlisting}
Inductive ty (B : Type) : Type :=
| Base : B -> ty B
| Fun : ty B -> ty B -> ty B.
\end{lstlisting}
Elements of \lstinline{B} are called \textbf{base types}.
Every inhabitant \lstinline{b : B} gives rise to a simple type \lstinline{Base b}
and if \lstinline{A1}, \lstinline{A2} are simple types
then so is \lstinline{Fun A1 A2}.
We write \lstinline{A1 ⟶ A2} for \lstinline{Fun A1 A2}.
\end{defi}

We need \emph{(variable) contexts} in order to type terms
that may contain free variables.
Conceptually,
a context is a list of variables with their respective types.
For instance,
\( [x_0 \hasType \atype_0; \ldots; x_n \hasType \atype_n] \)
is the context with variables \( x_0 \) of type
\( \atype_0 \), \ldots, \( x_n \) of type \( \atype_n \).
However,
we use nameless variables in our development,
so we do not keep track of their names.
Consequently,
a context is represented by a list of types.
Then %
we only consider the list \( [\atype_0, \ldots, \atype_n ] \).
However, we still need to refer to the free variables in terms.
In order to do so,
we represent them through indexing positions in the context.
For instance,
in the context \( [\atype_0; \dots; \atype_n] \)
we have \( n + 1 \) position indexes
\( 0, 1, \dots, n \), which we use as variables.

\begin{defi}[\coqident{Syntax.Signature.Contexts}{con}]
\label{def:ctx}
The type of \textbf{variable contexts} over a type \lstinline{B} is defined as follows.
\begin{lstlisting}
Inductive con (B : Type) : Type :=
| Empty : con B
| Extend : ty B -> con B -> con B.
\end{lstlisting}
We write \lstinline{∙} for \lstinline{Empty} and \lstinline{A ,,C} for \lstinline{Extend A C}.
\end{defi}
\begin{defi}[\coqident{Syntax.Signature.Contexts}{var}]
\label{def:vars}
We define the type \lstinline{var C A} of \textbf{variables} of type \lstinline{A} in context \lstinline{C} as
\begin{lstlisting}
Inductive var {B : Type} : con B -> ty B -> Type :=
| Vz : forall {C : con B} {A : ty B}, var (A ,, C) A
| Vs : forall {C : con B} {A1 A2 : ty B}, var C A2 -> var (A1 ,, C) A2.
\end{lstlisting}
\end{defi}

Let us consider an example of a context and some variables.
Suppose that we have a base type denoted by \lstinline{b}.
Then we can form the context \lstinline{Base b ,, Base b ⟶ Base b ,, Empty}.
In this context, we have two variables.
The first one, which is \lstinline{Vz}, has type \lstinline{Base b},
and the second variable, which is \lstinline{Vs Vz},
has type \lstinline{Base b ⟶ Base b},
The context that we discussed corresponds to \([x_0 : b ; x_1 : b \longrightarrow b]\).
The variable \lstinline{Vz} represents \(x_0\),
while \lstinline{Vs Vz} represents \(x_1\).

In \cref{def:terms-in-ctx} below we define the notion of
\emph{well-typed terms-in-context} which
consists of those expressions such that there is a typing derivation.
We use dependent types to ensure well-typedness of such expressions.
The type of terms depends on a simple type \lstinline{A : ty B}
(which represents the object-level type of the expression)
and context \lstinline{C : con B} that carries the types of all free variables in the term.
We also need to type function symbols.
Hence,
we require a type \lstinline{F : Type} of function symbols and
\lstinline{ar : F -> ty B},
which maps \lstinline{f : F} to a simple type \lstinline{ar f}.

\begin{defi}[\coqident{Syntax.Signature.Terms}{tm}]
\label{def:terms-in-ctx}
We define the type of \textbf{well-typed terms} as follows
\begin{lstlisting}
Inductive tm {B : Type} {F : Type} (ar : F -> ty B) (C : con B) : ty B -> Type :=
| BaseTm : forall (f : F), tm ar C (ar f)
| TmVar : forall {A : ty B}, var C A -> tm ar C A
| Lam : forall {A1 A2 : ty B}, tm ar (A1 ,, C) A2 -> tm ar C (A1 ⟶ A2)
| App : forall {A1 A2 : ty B}, tm ar C (A1 ⟶ A2) -> tm ar C A1 -> tm ar C A2.
\end{lstlisting}
For every function \lstinline{f : F} we have a term \lstinline{BaseTm f} of type \lstinline{ar f}.
Every variable \lstinline{v} gives rise to a term \lstinline{TmVar v}.
For \( \lambda \)-abstractions,
given a term \lstinline{s : tm ar (A1 ,, C) A2},
there is a term
\lstinline{λ s : tm ar C (A1 ⟶ A2)},
namely \lstinline{Lam s}.
The last constructor represents term application.
If we have a term \lstinline{s : tm ar C (A1 ⟶ A2)} and a term
\lstinline{t : tm ar C A1}, we get a term \lstinline{s · t : tm ar C A2},
which is defined to be \lstinline{App s t}.
\end{defi}

While it may be more cumbersome to write down terms using de Bruijn indices,
it does have several advantages.
Most importantly,
it eliminates the need for \( \alpha \)-equivalence,
so that determining equality between terms is reduced to a simple syntactic check.

Our notion of rewriting rules deviates slightly from the presentation
in~\cite{fuhs:kop:12}.
Mainly,
we do not impose the pattern restriction on the left-hand side of rules
nor that free variables on the right-hand side occur on the left-hand side.
This choice simplifies the formalization effort
because when defining a concrete TRS,
one does not need to check this particular condition.
Note that in {\isafor{}}~\cite{rene:ster:09} the same simplification is used

\begin{defi}[\coqident{Syntax.Signature}{rewriteRule}]
The type of \textbf{rewrite rules} is defined as follows:
\begin{lstlisting}
Record rewriteRule {B : Type} {F : Type} (ar : F -> ty B) :=
 make_rewrite {
    vars_of : con B ;
    tar_of : ty B ;
    lhs_of : tm ar vars_of tar_of ;
    rhs_of : tm ar vars_of tar_of }.
\end{lstlisting}
The context \lstinline{vars_of} carries the variables used in the rule,
and the type \lstinline{tar_of} is used to guarantee that both the
\lstinline{lhs_of} and \lstinline{rhs_of} are terms of the same type.
\end{defi}

\begin{defi}[\coqident{Syntax.Signature}{afs}]%
\label{def:afs}
The type of \textbf{algebraic functional systems} is defined as follows
\begin{lstlisting}
Record afs (B : Type) (F : Type) :=
make_afs { arity : F -> ty B ; list_of_rewriteRules : list (rewriteRule arity) }.
\end{lstlisting}
\end{defi}
As usual, every AFS induces a rewrite relation on the set of terms,
which we denote by \lstinline{∼>}.
The formal definition is found in \coqfile{Syntax.Signature}{RewritingSystem}.
The rewrite relation \lstinline{∼>} is defined to be the closure of the one-step relation
under transitivity and compatibility with the term constructors.
In Coq, we use an inductive type to define this relation.
Each rewrite step is represented by a constructor.
More specifically, we have a constructor for rewriting the left-hand and
the right-hand side of an application,
we have a constructor for \( \beta \)-reduction,
and we have a constructor for the rewrite rules of the AFS\@.

\begin{exa}[\coqident{Examples.Map}{map_afs}]%
\label{exa:map-afs}
Let us encode \( \rules_\map \) in Coq.
It is composed of two rules:
\( \map \app \aFuncVar \app \nil \arrz \nil \) and
\( \map \app \aFuncVar \app (\cons \app \avar \app \aListVar)
\arrz \cons \app (\aFuncVar \app \avar) \app (\map \app \aFuncVar \app \aListVar) \).
We start with base types.
\begin{lstlisting}
Inductive base_types := TBtype | TList.
Definition Btype : ty base_types := Base TBtype.
Definition List : ty base_types := Base TList.
\end{lstlisting}
The abbreviations \lstinline{Btype} and \lstinline{List} is to smoothen the usage of the base types.
There are three function symbols in this system:
\begin{lstlisting}
Inductive fun_symbols := TNil | TCons | TMap.
\end{lstlisting}
The arity function \lstinline{map_ar} maps each function symbol in \lstinline{fun_symbols} to its type.
\begin{lstlisting}
Definition map_ar f : ty base_types
   := match f with
      | TNil => List
      | TCons => Btype ⟶  List ⟶ List
      | TMap => (Btype ⟶  Btype) ⟶ List ⟶ List
      end.
\end{lstlisting}
So, \lstinline{TNil} is a list and given an inhabitant of \lstinline{Btype} and \lstinline{List},
the function symbol \lstinline{TCons} gives a \lstinline{List}.
Again we introduce some abbreviations to simplify the usage of the function symbols.
\begin{lstlisting}
Definition Nil {C} : tm map_ar C _ := BaseTm TNil.
Definition Cons {C} x xs : tm map_ar C _ := BaseTm TCons ·   x · xs.
Definition Map {C} f xs : tm map_ar C _ := BaseTm TMap ·   f · xs.
\end{lstlisting}
The first rule, \( \map \app \aFuncVar \app \nil \arrz \nil \),
is encoded as the following Coq construct:
\begin{lstlisting}
Program Definition map_nil :=
  make_rewrite
    (_ ,, ∙) _
    (let f := TmVar Vz in Map ·  f · Nil)
    Nil.
\end{lstlisting}
Notice that we only defined the \textit{pattern} of the first two arguments of
\lstinline{make_rewrite}, leaving the types in the context \lstinline{(_ ,, ∙)}
and the type of the rule unspecified.
Coq can fill in these holes automatically,
as long as we provide a context pattern of the correct length.
In this particular rewrite rule,
there is only one free variable.
As such,
the variable \lstinline{TmVar Vz} refers to the only variable in the context.
In addition,
we use iterated \lstinline{let}-statements to imitate variable names.
For every position in the context,
we introduce a variable in Coq,
which we use in the left- and right-hand sides of the rule.
This makes the rules more human-readable.
Indeed,
the lhs \( \map \app \app \aFuncVar \app \nil \) of this rule is represented
as \lstinline{Map · f · Nil} in code.
The second rule for \lstinline{map} is encoded following the same ideas.
\begin{lstlisting}
Program Definition map_cons :=
  make_rewrite
    (_ ,, _ ,, _ ,, ∙) _
    (let f := TmVar Vz in let x := TmVar (Vs Vz) in let xs := TmVar (Vs (Vs Vz)) in
    Map · f · (Cons · x · xs))
    (let f := TmVar Vz in let x := TmVar (Vs Vz) in let xs := TmVar (Vs (Vs Vz)) in
    Cons · (f · x) · (Map · f · xs)).
\end{lstlisting}
\end{exa}
Putting this all together, we obtain an AFS, which we call \lstinline{map_afs}.
\begin{lstlisting}
Definition map_afs := make_afs map_ar (map_nil :: map_cons :: nil).
\end{lstlisting}

\subsection{Termination}

Strong normalization is usually defined as the absence of infinite rewrite sequences.
Such a definition is sufficient in a classical setting
where the law of excluded middle holds.
However,
we work in a constructive setting,
and thus we are interested in a stronger definition.
Therefore, we need a constructive predicate, formulated positively,
which implies there are no infinite rewrite sequences.
This idea is captured by the following definition

\begin{defi}[\coqfile{Prelude.Relations}{WellfoundedRelation}]\label{def:is-wf}
The \textbf{well-foundedness predicate} for a relation \lstinline{R} is defined as follows
\begin{lstlisting}
Inductive isWf {X : Type} (R : X -> X -> Type) (x : X) : Prop :=
| acc : (forall (y : X), R x y -> isWf R y) -> isWf R x.
\end{lstlisting}
A relation is \textbf{well-founded} if the well-foundedness predicate holds for every element.
\begin{lstlisting}
Definition Wf {X : Type} (R : X -> X -> Type) :=
  forall (x : X), isWf R x.
\end{lstlisting}
\end{defi}

Note that this definition has been considered numerous times before, for example in~\cite{bertot:casteran:04} and in \COLORlib~\cite{blanqui:koprowski:11}.
An element \lstinline{x} is well-founded
if all \lstinline{y} such that \lstinline{R x y} are well-founded.
Note that if there is no \lstinline{y} such that \lstinline{R x y}, then \lstinline{x} is vacuously well-founded.
From the rewriting perspective,
this definition properly captures the notion of strong normalization.
Indeed,
a term \( \aterm \) is strongly normalizing iff
every \( \aterm' \) such that \( \aterm \) rewrites to \( \aterm'\)
is strongly normalizing.

Well-foundedness contradicts the existence of infinite rewrite sequences, even in a constructive setting.
As such, it indeed gives a stronger condition.

\begin{prop}[\coqident{Prelude.Relations.WellfoundedRelation}{no_infinite_chain}]
If \( R \) is well-founded,
then there is no infinite sequence
\( \aterm_0, \aterm_1, \ldots \) such that
\( R(\aterm_n, \aterm_{n + 1}) \), for all \( n \).
\end{prop}

Next, we define strong normalization using well-founded predicates.

\begin{defi}[\coqident{Syntax.StrongNormalization.SN}{is_SN}]
An algebraic functional system is \textbf{strongly normalizing} if for every context \lstinline{C}  and every type \lstinline{A} the rewrite relation for terms of type \lstinline{A} in context \lstinline{C} is well-founded.
We formalize that as follows:
\begin{lstlisting}
Definition isSN {B F : Type} (X : afs B F) : Prop :=
  forall (C : con B) (A : ty B), Wf (fun (t1 t2 : tm X C A) => t1 ∼> t2).
\end{lstlisting}
\end{defi}

\section{Higher-Order Interpretation Method}\label{sec:weak-int}

In this section,
we formalize the method of weakly monotonic algebras for algebraic functional systems.
We proceed by providing type-theoretic semantics for the syntactic constructions
introduced in the last section and a sufficient condition for which
such semantics can be used to establish strong normalization.

\subsection{Interpreting types and terms}
In weakly monotonic algebras, types are interpreted as sets along with a well-founded
ordering and a quasi-ordering~\cite{pol:96,fuhs:kop:12}.
For that reason, we start by defining \textit{compatible relations}.
Intuitively, these are the domain for our semantics.

\begin{defi}[\coqfile{Prelude.Orders}{CompatibleRelation}]
\textbf{Compatible relations} are defined as follows
\begin{lstlisting}
Record CompatRel := {
  carrier :> Type ;
  gt : carrier -> carrier -> Prop ;
  ge : carrier -> carrier -> Prop }.
\end{lstlisting}
We write \lstinline{x > y} and \lstinline{x >= y} for \lstinline{gt x y} and \lstinline{ge x y} respectively.
\end{defi}
The record \lstinline{CompatRel} consists of the data needed to express
compatibility between \lstinline{>}~and~\lstinline{>=}.
The conditions it needs to satisfy, are in the type class
\lstinline{isCompatRel}, defined below.
\begin{lstlisting}
Class isCompatRel (X : CompatRel) := {
  gt_trans : forall {x y z : X}, x > y -> y > z -> x > z ;
  ge_trans : forall {x y z : X}, x >= y -> y >= z -> x >= z ;
  ge_refl : forall (x : X), x >= x ;
  compat : forall {x y : X}, x > y -> x >= y ;
  ge_gt : forall {x y z : X}, x >= y -> y > z -> x > z ;
  gt_ge : forall {x y z : X}, x > y -> y >= z -> x > z }.
\end{lstlisting}

Note that the field \lstinline{gt_trans} in \lstinline{isCompatRel} follows from \lstinline{compat} and \lstinline{ge_gt}.
The type \lstinline{nat} of natural numbers with the usual orders is a first example of data
that satisfies \lstinline{isCompatRel}.
We denote this one by \lstinline{nat_CompatRel}.
This type class essentially models
the notion of extended well-founded set introduced in~\cite{kop:vale:21}.
An \textbf{extended well-founded set} is a set together with compatible
orders \( \costGt, \costGe \) such that \( \costGt \) is well-founded
and \( \costGe \) is a quasi-ordering.
This compatibility requirement corresponds to the axiom \lstinline{compat}
in the type class \lstinline{isCompatRel}.
However,
since we do not require \( > \) to be well-founded in this definition,
we instead call it a compatible relation.
More specifically,
\lstinline{X} is a compatible relation if it is of type \lstinline{CompatRel}
and satisfies the constraints in the type class \lstinline{isCompatRel}.

In order to interpret simple types (\cref{def:stypes}),
we start by fixing a type \lstinline{B : Type} of base types and
an interpretation {\lstinline{semB : B -> CompatRel}}
such that each \lstinline{semB b} is a compatible relation.
Whenever {\lstinline{semB}} satisfies such property
we call it an \textbf{interpretation key} for {\lstinline{B}}.
We interpret arrow types as functional compatible relations, i.e.,
compatible relations such that the inhabitants of their carrier are functional.
The class of functionals we are interested in is that of \textit{weakly-monotone maps}.

\begin{defi}[\coqfile{Prelude.Orders}{MonotonicMaps}]
\textbf{Weakly monotone maps} are defined as follows
\begin{lstlisting}
Class weakMonotone {X Y : CompatRel} (f : X -> Y) :=
  map_ge : forall (x y : X), x >= y -> f x >= f y.

Record weakMonotoneMap (X Y : CompatRel) :=
  make_monotone {
    fun_carrier :> X -> Y ;
    is_weak_monotone : weakMonotone fun_carrier }.
\end{lstlisting}
The class \lstinline{weakMonotone} says when a function is weakly monotonic,
and an inhabitant of the record \lstinline{weakMonotoneMap}
consists of a function together with proof of its weak monotonicity.
Then we define \lstinline{fun_CompatRel}
which is of type \lstinline{CompatRel}
and represents the
\textbf{functional compatible relations} from \lstinline{X} to \lstinline{Y}.
It is defined as follows:
\begin{lstlisting}
Definition fun_CompatRel (X Y : CompatRel) : CompatRel :={|
  carrier := weakMonotoneMap X Y ;
  gt f g := forall (x : X), f x > g x ;
  ge f g := forall (x : X), f x >= g x |}.
\end{lstlisting}
\end{defi}
In what follows,
we write \lstinline{X →wm Y} for \lstinline{fun_CompatRel X Y}.
The semantics for a type
is parametrized by an interpretation key \lstinline{semB}.
It is defined as follows:
\begin{defi}[\coqident{Interpretation.OrderInterpretation}{sem_Ty}]%
\label{def:sem_Ty}
Assume {\lstinline{A : ty B}}
and {\lstinline{semB}} is an interpretation key for {\lstinline{B}}.
Then
\begin{lstlisting}
Fixpoint sem_Ty (A : ty B) : CompatRel :=
    match A with
    | Base b    => semB b
    | A1 → A2  => sem_Ty A1 →wm sem_Ty A2
    end.
\end{lstlisting}
\end{defi}
We also show how to interpret contexts,
and to do so,
we need to interpret the empty context and context extension.
For those, we define the unit and product of compatible relations.

\begin{defi}[\coqfile{Prelude.Orders}{Examples}]
The \textbf{unit} and \textbf{product} compatible relations:\\
\begin{tabular}{ll}
  \begin{lstlisting}
    Definition unit_CompatRel :
    CompatRel := {|
      carrier := unit ;
      gt _ _ := False ;
      ge _ _ := True |}.
    \end{lstlisting}
    &
    \begin{lstlisting}
      Definition prod_CompatRel (X Y : CompatRel) :
      CompatRel := {|
        carrier := X * Y ;
        gt x y := fst x > fst y /\ snd x > snd y ;
        ge x y := fst x >= fst y /\ snd x >= snd y |}.
    \end{lstlisting}
\end{tabular}
\end{defi}

Note that \lstinline{unit_CompatRel} is the compatible relation on the
type with only one element, for which the ordering is trivial.
In addition, \lstinline{prod_CompatRel} is the compatible relation on the
product, for which we compare elements coordinate-wise.
We write \lstinline{X * Y} for \lstinline{prod_CompatRel X Y}.

\begin{defi}[\coqident{Interpretation.OrderInterpretation}{sem_Con}]%
\label{def:sem-ctx}
Contexts are interpreted as follows
\begin{lstlisting}
Fixpoint sem_Con (C : con B) : CompatRel :=
   match C with
   | ∙        => unit_CompatRel
   | A ,, C  => sem_Ty A * sem_Con C
   end.
\end{lstlisting}
\end{defi}

Next, we give semantics to variables and terms.
The approach we use here is slightly different from what is usually done
in higher-order rewriting.
In~\cite{fuhs:kop:12,kop:vale:21,pol:96}, for instance,
context information is lifted to the meta-level
and variables are interpreted using the notion of valuations.
In contrast,
in our setting, the typing context lives at the syntactic level
and variables are interpreted as weakly monotonic functions.
Consequently,
to every term \lstinline{t : tm C A},
we assign a map from \lstinline{sem_Con C} to \lstinline{sem_Ty A}.
In the remainder, we need the following weakly monotonic functions.

\begin{defi}[\coqfile{Prelude.Orders}{Examples}]\label{def:combinators-wm}
We define the following weakly monotonic functions.
\begin{itemize}
    \item Given \lstinline{y : Y},
    we write \lstinline{const_wm y : X →wm y}
    for the constant function.
    \item Given \lstinline{f : X →wm Y} and \lstinline{g : Y →wm Z},
    we define \lstinline{g ∘wm f : X →wm Z} to be their composition.
    \item We have the first projection \lstinline{fst_wm : X * Y →wm X},
    which sends a pair \lstinline{(x , y)} to \lstinline{x},
    and the second projection \lstinline{snd_wm : X * Y →wm Y},
    which sends \lstinline{(x , y)} to \lstinline{y}.
    \item Given \lstinline{f : X →wm Y} and \lstinline{g : X →wm Z},
    we have a function \lstinline{⟨ f , g ⟩ : X →wm (Y * Z)}.
    For \lstinline{x : X}, we define \lstinline{⟨ f , g ⟩ x} to be \lstinline{(f x , g x)}.
    \item Given \lstinline{f : Y * X →wm Z},
    we get \lstinline{λwm f : X →wm (Y →wm Z)}.
    For every \lstinline{x : X} and \lstinline{y : Y},
    we define \lstinline{λwm f y x} to be \lstinline{f (y , x)}.
    \item Given \lstinline{f : X →wm (Y →wm Z)} and \lstinline{x : X →wm Y},
    we obtain \lstinline{f ·wm x : X →wm Z},
    which sends every \lstinline{a : X}
    to \lstinline{f a (x a)}.
    \item Given \lstinline{x : X},
    we have a weakly monotonic function
    \lstinline{apply_el_wm x : (X →wm Y) →wm Y}
    which sends \lstinline{f : X →wm Y}
    to \lstinline{f x}.
\end{itemize}
\end{defi}

Recall that variables are represented by positions in a context which in turn
is interpreted as a weakly monotonic product~(\cref{def:sem-ctx}).
This allows us to interpret the variable at a position in a context
as the corresponding interpretation of the type in that position.

\begin{defi}[\coqident{Interpretation.OrderInterpretation}{sem_Var}]
We interpret variables with the following function
\begin{lstlisting}
Fixpoint sem_Var {C : con B} {A : ty B} (v : var C A) : sem_Con C →wm sem_Ty A
  := match v with
    | Vz => fst_wm
    | Vs v => sem_Var v ∘wm snd_wm
    end.
\end{lstlisting}
\end{defi}

We need the following data in order to provide semantics to terms.
An arity function \lstinline{ar : F -> ty B}, together with its interpretation
\lstinline{semF : forall (f : F), sem_Ty (ar f)},
and an \textit{application operator} given by
\begin{lstlisting}
semApp : forall (A1 A2 : ty B), (sem_Ty A1 →wm sem_Ty A2) * sem_Ty A1 →wm sem_Ty A2
\end{lstlisting}
to interpret term application.

\begin{rem}
A first, but incorrect, guess to interpret application would have been by
interpreting the application of \lstinline{f : sem_Ty A1 →wm sem_Ty A2}
to \lstinline{x : sem_Ty A1} by \lstinline{f x}.
However, there is a significant disadvantage of this interpretation.
Ultimately, we want to deduce strong normalization from the interpretation,
and the main idea is that if we have a rewrite \lstinline{x ∼> x'},
then we have \lstinline{semTm x > semTm x'}.
This requirement would not be satisfied if we interpret application of
our terms as actual applications as functions.
Indeed, if we have \lstinline{x < x'},
then one is not guaranteed that we also have \lstinline{f x < f x'},
because \lstinline{f} is only weakly monotone.

There are two ways to deal with this.
One way is by interpreting function types as strictly monotonic maps~\cite{kop:vale:21}.
In this approach, this interpretation of application is valid.
However, it comes at a price, because the interpretation of lambda abstraction
becomes more difficult.

Another approach, which we use here, is also used~in~\cite{fuhs:kop:12}.
We add a parameter to our interpretation method,
namely \lstinline{semApp},
which abstractly represents the interpretation of application.
To deduce strong normalization in this setting,
we add requirements about \lstinline{semApp} in \cref{sec:termination-model}.
As a result, in concrete instantiations of this method,
we need to provide an actual definition for \lstinline{semApp}.
We see this in \cref{sec:poly-int}.
\end{rem}

\begin{defi}[\coqident{Interpretation.OrderInterpretation}{sem_Tm}]\label{def:int-terms}
Given a function \lstinline{semF : forall (f : F), sem_Ty (ar f)},
the semantics of terms is given by
\begin{lstlisting}
Fixpoint sem_Tm {C : con B} {A : ty B} (t : tm ar C A) : sem_Con C →wm sem_Ty A :=
  match t with
  | BaseTm f  => const_wm (semF f)
  | TmVar v   => sem_Var v
  | λ f       => λwm (sem_Tm f)
  | f · t     => semApp _ _ ∘wm ⟨ sem_Tm f , sem_Tm t ⟩
  end.
\end{lstlisting}
\end{defi}

Notice that we could have chosen a fixed way of interpreting application.
We follow the same approach used by Fuhs and Kop~\cite{fuhs:kop:12}
in our formalization and leave \lstinline{semApp} abstract.
This choice is essential if we want to use the interpretation method for both
\textit{rule removal} and the \textit{dependency pair} approach.
See~\cite[Chapters 4 and 6]{kop:12} for more detail.

\subsection{Termination Models for AFSs}
\label{sec:termination-model}
Now we have set up everything that is necessary to define the main
notion of this section: \emph{termination models}.
From a termination model of an algebraic functional system, one obtains an
interpretation of the types and terms.
In addition, every rewrite rule is `satisfied' in this interpretation.

\begin{defi}[\coqident{Interpretation.OrderInterpretation}{Interpretation}]%
\label{def:term-model}
Let \( \rules \) be an algebraic functional system with base type
\lstinline{B} and function symbols \lstinline{F}.
A \textbf{termination model} of \( \rules \) consists of
\begin{itemize}
	\item an interpretation key \lstinline{semB};
	\item a function \lstinline{semF : forall (f : F), sem_Ty (ar f)};
	\item a function
	\begin{lstlisting}
	semApp : forall (A1 A2 : ty B), (sem_Ty A1 →wm sem_Ty A2) * sem_Ty A1 →wm sem_Ty A2
	\end{lstlisting}
\end{itemize}
such that the following axioms are satisfied
\begin{itemize}
	\item each \lstinline{semB b} is well-founded and inhabited;
	\item if \lstinline{f > f'}, then \lstinline{semApp _ _ (f , x) > semApp _ _ (f' , x)};
	\item if \lstinline{x > x'}, then \lstinline{semApp _ _ (f , x) > semApp _ _ (f , x')};
	\item we have \lstinline{semApp _ _ (f , x) >= f x} for all \lstinline{f} and \lstinline{x};
	\item for every rewrite rule \lstinline{r},
	substitution \lstinline{s},
	and element \lstinline{x}, we have
	\begin{lstlisting}
	semTm (lhs r [ s ]) x > semTm (rhs r [ s ]) x.
	\end{lstlisting}
\end{itemize}
\end{defi}

Whereas the left-hand side of every rewrite rule is greater than its right-hand side,
this does not hold for \(\beta \)-reduction in our interpretations.
Since rewrite sequences can contain both rewrite rules and \(\beta \)-reduction,
such sequences are not guaranteed to strictly decrease.
As such, we need more to actually conclude strong normalization, and we follow the method used by Kop~\cite{kop:12}.
More specifically, Kop uses \emph{rule removal} to show that
strong normalization follows from the strong normalization of $\beta$-reduction,
which is a famous theorem proven by Tait~\cite{taitmethod}.
The strong normalization of $\beta$-reduction has been formalized numerous times
and an overview can be found in~\cite{AbelAHPMSS19}.
Now we deduce the main theorem of this section.

\begin{thrm}[\coqident{Interpretation.OrderInterpretation}{afs_is_SN_from_Interpretation}]%
\label{thm:sn-interpretation}
Let \( \rules \) be an algebraic functional system.
If we have a termination model of \( X \),
then \( X \) is strongly normalizing.
\end{thrm}

\section{The Higher-Order Polynomial Method}

\subsection{Polynomials}
In this section,
we instantiate the material of \cref{sec:weak-int} to a concrete instance,
namely \emph{the polynomial method}~\cite{fuhs:kop:12}.
For that reason, we define the notation of \emph{higher-order polynomial}.

\begin{defi}[\coqfile{TerminationTechniques.PolynomialMethod}{Polynomial}]
We define the type \lstinline{base_poly} of \textbf{base polynomials} and
\lstinline{poly} of \textbf{higher-order polynomials}
by mutual induction as follows:\\
\begin{tabular}{@{}ll@{}}
\begin{lstlisting}
  Inductive base_poly {B : Type}
      : con B -> Type :=
  | P_const : forall {C : con B},
      nat -> base_poly C
  | P_plus : forall {C : con B},
      (P1 P2 : base_poly C) -> base_poly C
  | P_mult : forall {C : con B},
      (P1 P2 : base_poly C)-> base_poly C
  | from_poly : forall {C : con B} {b : B},
      poly C (Base b) -> base_poly C
\end{lstlisting}
&
\begin{lstlisting}
  with poly {B : Type} : con B → ty B → Type :=
  | P_base : forall {C : con B} {b : B},
    base_poly C → poly C (Base b)
  | P_var : forall {C : con B} {A : ty B},
    var C A → poly C A
  | P_app : forall {C : con B} {A₁ A₂ : ty B},
    poly C (A₁ ⟶ A₂)
    → poly C A₁
    → poly C A₂
  | P_lam : forall {C : con B} {A₁ A₂ : ty B},
    poly (A₁ ,, C) A₂ → poly C (A₁ ⟶ A₂).
\end{lstlisting}
\end{tabular}
\end{defi}
We can make expressions of base polynomials using \lstinline{P_const} (constants), \lstinline{P_plus} (addition), and \lstinline{P_mult} (multiplication).
In addition, \lstinline{from_poly} takes an inhabitant of \lstinline{poly C (Base b)} and returns a base polynomial in context \lstinline{C}.
Using \lstinline{P_base}, we can turn a base polynomial into a polynomial of any base type.
The constructors, \lstinline{P_var}, \lstinline{P_app}, and \lstinline{P_lam}, are remniscent of the simply typed lambda calculus.
We get variables from \lstinline{P_var},
$\lambda$-abstraction from \lstinline{P_lam},
and application from \lstinline{P_app}.
Note that combining \lstinline{from_poly} and \lstinline{P_var},
we can use variables in base polynomials.

Let us make some remarks about the design choices we made and how they affected the definition of polynomials.
One of our requirements is that we are able to add and multiply polynomials on different base types.
This is frequently used in actual examples, such as \cref{exa:map-poly}.
Function symbols might use arguments from different base types,
and we would like to use both of them in polynomial expressions.

One possibility would have been to only work with the type \lstinline{poly} and to add a constructor
\begin{lstlisting}
P_plus : forall {C : con B} (b1 b2 : B),
         poly C (Base b1) -> poly C (Base b2) -> poly C (Base b1)
\end{lstlisting}
However, we refrained from doing so: if we were to use \lstinline{P_const}, then the elaborator would be unable to determine the actual type if we do not tell the base type explicitly.
Instead, we used a type of base polynomials that does not depend on the actual base type.
This is the role of \lstinline{base_poly}, which only depends on the variables being used.
We can freely add and multiply inhabitants of \lstinline{base_poly}, and if we were to use a constant, then we do not explicitly need to mention the base type.
In addition, we are able to transfer between \lstinline{base_poly} and \lstinline{poly C (Base b)}, and that is what \lstinline{P_base} and \lstinline{from_poly} enable us to do.

Note that our definition of higher-order polynomials is rather similar to the one given by Fuhs and Kop~\cite[Definition 4.1]{fuhs:kop:12}.
They define a set $\pol{X}$, which consists of polynomial expressions, and for every type $A$ a set $\polTy{X}{A}$.
The set $\polTy{X}{A}$ is defined by recursion: for base type, it is the set of polynomials over $X$ and for function types $A_1 \arrfunc A_2$, it consists of expressions $\Lambda (y : A_1). P$ where $P$ is a polynomial of type $A_2$ using an extra variable $y : A_1$.
Our \lstinline{base_poly C} and \lstinline{poly C A} correspond to $\pol{X}$ and $\polTy{X}{A}$ respectively.
However, there are some differences.
First of all, Fuhs and Kop require variables to be fully applied, whereas we permit partially applied variables.
Secondly, Fuhs and Kop define polynomials in such a way that for every two base types $b_1, b_2$ the types $\polTy{X}{b_1}$ and $\polTy{X}{b_2}$ are equal.
This is not the case in our definition: instead we use constructors \lstinline{from_poly} and \lstinline{P_base} to relate \lstinline{base_poly C} and \lstinline{poly C (Base b)}.

In the polynomial method,
the interpretation key sends every base type to \lstinline{nat_CompatRel},
and in what follows, we write \lstinline{⟦ C ⟧con} and \lstinline{⟦ A ⟧ty}
for the interpretation of contexts and types respectively.
Note that every polynomial \lstinline{P : poly C A} gives rise to a weakly monotonic function \lstinline{sem_poly P : ⟦ C ⟧con →wm ⟦ A ⟧ty}
and that every base polynomial \lstinline{P : base_poly C} gives rise to \lstinline{sem_base_poly P : ⟦ C ⟧con →wm nat_CompatRel}.
These two functions are defined using mutual recursion
and every constructor is interpreted in the expected way: \coqident{TerminationTechniques.PolynomialMethod.Polynomial}{sem_poly}.

In order to actually use \lstinline{base_poly C} and \lstinline{poly C A},
we provide convenient notations for operations on polynomials.
More concretely, we define notations \lstinline{+},
\lstinline{*}, and \lstinline{·P} that represent addition,
multiplication, and application respectively.
These operations must be overloaded since we need to be able to add polynomials of different types.
To do so, we similarly use type classes to \textsf{MathClasses}~\cite{spitters:weegen:11}.
For details, we refer the reader to the formalization.

\begin{exa}[\coqident{Examples.Map}{map_fun_poly}]\label{exa:map-poly}
We continue with \cref{exa:map-afs} and provide a polynomial interpretation to
the system \lstinline{map_afs} as follows:
\begin{lstlisting}
Definition map_fun_poly fn_symbols : poly ∙ (arity trs fn_symbols) :=
   match fn_symbols with
   | Tnil => to_Poly (P_const 3)
   | Tcons  => λP λP let y1 := P_var Vz in
     to_Poly (P_const 3 + P_const 2 * y1)
   | Tmap  =>  λP let y0 := P_var (Vs Vz) in λP let G1 := P_var Vz in
     to_Poly (P_const 3 * y0 + P_const 3 * y0 * (G1 ·P (y0)))
   end.
\end{lstlisting}
Informally, the interpretation of \lstinline{nil} is the constant 3.
The interpretation of \lstinline{cons} is the function that sends $y_1 : \mathbb{N}$ to $3 + 2 y_1$,
and \lstinline{map} is interpreted as the function that sends $y_0 : \mathbb{N}$
and $G_1 : \mathbb{N} \arrfuncwm \mathbb{N}$
to $3 y_0 + 3 y_0 G_1(y_0)$.
\end{exa}

\subsection{Polynomial Interpretation}\label{sec:poly-int}
Using polynomials, we deduce strong normalization under certain circumstances using \cref{thm:sn-interpretation}.
Suppose that for all function symbols \lstinline{f} we have a polynomial \lstinline{J : poly ∙ (arity X f)},
and now we need to provide the interpretation for application.
Following {Fuhs and Kop}~\cite{fuhs:kop:12},
we use a general method to interpret application.
We start by constructing a minimal element in the interpretation of every type.

\begin{defi}[\coqident{TerminationTechniques.PolynomialMethod.Polynomial}{min_el_ty}]
For every simple type \lstinline{A} we define a minimal element of \lstinline{⟦ A ⟧ty} as follows
\begin{lstlisting}
Fixpoint min_el_ty (A : ty B) : minimal_element ⟦ A ⟧ty
   := match A with
      | Base _ => nat_minimal_element
      | A1 ⟶ A2 => min_el_fun_space (min_el_ty A2)
      end.
\end{lstlisting}
Here \lstinline{nat_minimal_element} is defined to be 0,
and \lstinline{min_el_fun_space (min_el_ty A2)} is the constant function on \lstinline{(min_el_ty A2)}.
\end{defi}

In order to define the semantics of application,
we need several operations involving \lstinline{⟦ A ⟧ty}.
First, we consider \emph{lower value functions}.

\begin{defi}[\coqident{TerminationTechniques.PolynomialMethod.Polynomial}{lvf}]
We define the \textbf{lower value function} as follows
\begin{lstlisting}
Fixpoint lvf {A : ty B} : ⟦ A ⟧ty →wm nat_CompatRel :=
  match A with
    | Base _ => id_wm
    | A1 ⟶ A2 => lvf ∘wm apply_el_wm (min_el_ty A1)
  end.
\end{lstlisting}
\end{defi}
Note that we construct \lstinline{lvf} directly as a weakly monotonic function.
In addition, we reuse the combinators defined in \cref{def:combinators-wm}.
As such, we do not need to prove separately that this function is monotonic.

In Kop and Fuhs~\cite{fuhs:kop:12}, this definition is written down in a different, but equivalent, way.
Instead of defining $\lvf{\atype}$ recursively, they look at full applications,
which would be more complicated in our setting.
More specifically, since we are working with simple types, we must have that $\atype = \atype_1 \arrfunc \ldots \arrfunc \atype_n \arrfunc \btype$.
Then they define
\(
\lvf{\atype}(f) \defeq f(\minEl{\atype_1}, \ldots, \minEl{\atype_n}),
\)
where $\minEl{\atype}$ is the minimum element of the interpretation of $\atype$.
Next, we define two addition operations on \lstinline{⟦ A ⟧ty}.

\begin{defi}[\coqident{TerminationTechniques.PolynomialMethod.Polynomial}{plus_ty_nat}]\label{def:plus_ty_nat}
Addition of natural numbers and elements on \lstinline{⟦ A ⟧ty} is defined as follows
\begin{lstlisting}[mathescape=true]
Fixpoint plus_ty_nat {A : ty B} : ⟦ A ⟧ty * nat_CompatRel →wm ⟦ A ⟧ty
   := match A with
      | Base _ => plus_wm
      | A1 ⟶ A2 =>
        let f := fst_wm ∘wm snd_wm in
        let x := fst_wm in
        let n := snd_wm ∘wm snd_wm in
        $\lambda$wm (plus_ty_nat ∘wm ⟨ f ·wm x , n ⟩)
      end.
\end{lstlisting}
\end{defi}

The function \lstinline{plus_ty_nat} allows us to add arbitrary natural
numbers to elements of the interpretation of types.
Note that there are two cases in \cref{def:plus_ty_nat}.
First of all, the type \lstinline{A} could be a base type.
In that case, we are adding two natural numbers, and we use the usual addition operation.
In the second case, we are working with a functional type \lstinline{A1 ⟶ A2}.
The resulting function is defined using pointwise addition with the relevant natural number.
Now we have everything in place to define the interpretation of application.

\begin{defi}[\coqident{TerminationTechniques.PolynomialMethod.Polynomial}{p_app}]\label{def:interpret-app}
Application is interpreted as the following function
\begin{lstlisting}
Definition p_app {A1 A2 : ty B}
  : ⟦ A1 ⟶ A2 ⟧ty * ⟦ A1 ⟧ty →wm ⟦ A2 ⟧ty
  := let f := fst_wm in
     let x := snd_wm in
     plus_ty_nat ∘wm ⟨ f ·wm x , lvf ∘wm x ⟩.
\end{lstlisting}
\end{defi}

If both \lstinline{A1} and \lstinline{A2} are base types, then \lstinline{p_app (f , x)}
reduces to \lstinline{f x + x}.
Note that \lstinline{p_app} satisfies the requirements from \cref{thm:sn-interpretation}.
Hence, we obtain the following.

\begin{thrm}[\coqident{TerminationTechniques.PolynomialMethod.Polynomial}{poly_Interpretation}]%
\label{thm:poly-sn}
Let \( \rules \) be an AFS\@.
Suppose that for every function symbol
\textup{\lstinline{f}} we have a polynomial \textup{\lstinline{p_fun_sym f}}
such that
for all rewrite rules \textup{\lstinline{l ∼> r}} in \( \rules \)
we have
\textup{\lstinline{semTm l x > semTm r x}}
for all \textup{\lstinline{x}}.
Then \( \rules \) has a termination model.
\end{thrm}

\subsection{Constraint Solving Tactic}\label{sec:poly-notations}

Notice that in order to formally verify a proof of termination of a system
using \cref{thm:poly-sn},
we need to provide a polynomial interpretation and show that
\( \interpret{\ell} \costGt \interpret{r} \) holds
for all rules \( \ell \arrz r \).
This will introduce inequality proof goals into the Coq context that
must be solved.

\begin{exa}\label{exa:poly_ex}
  Let us consider a concrete example.
  We use the polynomials given in \cref{exa:map-poly}
  to show strong normalization of \cref{exa:map-afs}.
  This example introduces two inequalities, one for each rule.
  Let \( G_0 : \Nat \arrfuncwm \Nat \) be weakly monotonic.
  For rule \lstinline{map_nil},
  we need to prove that for all %
  \( G_0 \),
  the constraint \( 12 + G_0(0) + 9 G_0(3) \costGt 3 \) holds.
  For the second rule, \lstinline{map_cons}, the constraint is:
  \(
    12 + 4 y_0 + 12 y_1 + G_0(0) + (3 y_0 + 9 y_1 + 9) G_0(3 + y_0 + 3 y_1)
    \costGt
    3 + y_0 + 12 y_1 + 3 G_0(0) + G_0(y_0) + 9 y_1 G_0(y_1)
  \),
  for all \( y_0, y_1 \in \Nat \)
  and \( G_0 \).
\end{exa}

Finding witnesses for such inequalities is tedious,
and we would like to automate this task.
For that reason,
we developed a tactic
(\coqident{TerminationTechniques.PolynomialMethod.Polynomial}{solve_poly})
that automatically solves
the inequalities coming from \cref{thm:poly-sn}.
Essentially,
this tactic tries to mimic how one would solve those goals in
a pen-and-paper proof,
and the same method is used by \wanda{}.

\begin{exa}
We show how to solve the constraint arising from \lstinline{map_cons}
mentioned in \cref{exa:poly_ex}.
The first step is to find matching terms on both sides of the inequality
and subtract them.
In our example,
\( 3 + y_0 + 12 y_1 + G_0(0) \) occurs on both sides,
and after subtraction,
we obtain the following constraint:
\[
  9 + 3 y_0 + 9 y_1 + (3 y_0 + 9 y_1 + 9) G_0(3 + y_0 + 3 y_1)
  \costGt
  2 G_0(0) + G_0(y_0) + 9 y_1 G_0(y_1).
\]
The second step is combining the arguments for the higher-order variable \( G_0 \)
use its monotonicity.
Note that each of \( 0 \), \( y_0 \),
and \( y_1 \) is lesser than or equal to \( 3 + y_0 + 3 y_1 \),
because they are natural numbers.
Since \( G_0 \) is weakly monotonic,
we get
\[
2 G_0(0) + G_0(y_0) + 9 y_1 G_0(y_1) \leq (9 y_1 + 3) G_0(3 + y_0 + 3 y_1).
\]
Now we can simplify our original constraint to
\[
  9 + 3 y_0 + 9 y_1 + (3 y_0 + 9 y_1 + 9) G_0(3 + y_0 + 3 y_1)
  \costGt
  (9 y_1 + 3) G_0(3 + y_0 + 3 y_1).
\]
Since $3 y_0 + 9 y_1 + 9 \geq 9 y_1 + 3$, we have
\[
  (3 y_0 + 9 y_1 + 9) G_0(3 + y_0 + 3 y_1)
  \geq
  (9 y_1 + 3) G_0(3 + y_0 + 3 y_1).
\]
This is sufficient to conclude that the constraints
for \lstinline{map_cons} are satisfied.
\end{exa}
The tactic \lstinline{solve_poly} (\coqident{TerminationTechniques.PolynomialMethod.PolynomialTactics}{solve_poly}) follows the steps described above.
Note that we use the tactic \lstinline{nia},
which is a tactic in Coq that can solve inequalities and equations in nonlinear integer arithmetic.
More specifically, \lstinline{solve_poly} works as follows:
\begin{itemize}
	\item First, we generate a goal for every rewrite rule, and we destruct the
	assumptions so that each variable in the context is either a natural number or
	a function.
	\item For every variable $f$ that has a function type,
	we look for pair $(x, y)$ such that $f(x)$ on the left hand side
	and $f(y)$ occurs on the right-hand side.
	We try using \lstinline{nia} whether we can prove $x < y$ from our
	assumptions.
	If so, we add $x < y$ to the assumptions,
	and otherwise, we continue.
	\item The resulting goals with the extra assumptions are solved using \lstinline{nia}.
\end{itemize}

Note that \lstinline{solve_poly} is not complete, because \lstinline{nia} is incomplete.
As such, if a proof using this tactic is accepted by Coq, then that proof is correct.
However, if the proof is not accepted, then it does not have to be the case that
the proof is false.
With the material discussed in this section,
we can write down the polynomials given in \cref{exa:map-poly},
and the tactic is able to verify strong normalization.

\section{Generating Proof Scripts}\label{sec:proof-scripts}

In this section,
we discuss the practical aspects of our verification framework.
In principle
one can manually encode rewrite systems as Coq files and use the formalization
we provide to verify their own termination proofs.
However,
this is cumbersome to do.
Indeed,
in \cref{exa:map-afs} we used abbreviations to make the formal description
of \( \rules_\map \) more readable.
A rewrite system with many more rules would be difficult to encode manually.
Additionally,
to formally establish termination we also need to encode proofs.
We did this in \cref{exa:map-poly}.
The full formal encoding of \( \rules_\map \) and its termination proof
is found in the file \coqfile{Examples}{Map}.

\subsection{Proof traces for polynomial interpretation}
This difficulty of manual encoding motivates the usage of proof traces.
A proof trace is a human-friendly encoding of a TRS and the essential
information needed to reconstruct the termination proof as a Coq script.
Let us again consider \( \rules_\map \) as an example.
The proof trace for this system starts with \texttt{YES} to signal that
we have a termination proof for it.
Then we have a list encoding the signature and the rules of the system.
\begin{verbatim}
YES
Signature: [
  cons : a -> list -> list ;
  map : list -> (a -> a) -> list ;
  nil : list
]
Rules: [
  map nil F => nil ;
  map (cons X Y) G => cons (G X) (map Y G)
]
\end{verbatim}
Notice that the free variables in the rules do not need to be declared
nor their typing information provided.
Coq can infer this information automatically.
The last section of the proof trace describes the interpretation of each
function symbol in the signature.

\begin{verbatim}
Interpretation: [
  J(cons) = Lam[y0;y1].3 + 2*y1;
  J(map)  = Lam[y0;G1].3*y0 + 3*y0 * G1(y0);
  J(nil)  = 3
]
\end{verbatim}

We can fully reconstruct a formal proof of termination for \( \rules_\map \),
which uses \cref{thm:poly-sn},
with the information provided in the proof trace above.
The full description of proof traces can be found in~\cite{vale:weide:22},
the API for \onijn{}\@.
Proof traces are not Coq files.
So we need to further compile them
into a proper Coq script.
The schematics in~\cref{fig:onij-nijn-schm} describe the steps necessary for it.
We use \onijn{} to compile proof traces to Coq script.
It is invoked as follows:
\begin{lstlisting}[language=bash, style=custombash]
  (onijn) <path/to/proof/trace.onijn> (-o) <path/to/proof/script.v>
\end{lstlisting}
Here, the first argument is the file path to a proof trace file and the
\lstinline[language=bash, style=custombash]{(-o)} option requires the file path to the resulting Coq script.
The resulting Coq script can be verified by \nijn{} as follows:
\begin{lstlisting}[language=bash, style=custombash]
  (coqc) <path/to/proof/script.v>
\end{lstlisting}
Instructions on how to locally install \onijn{}/\nijn{}
can be found at~\cite{vale:weide:22}.

\subsection{Verifying Wanda's Polynomial Interpretations}

It is worth noticing that
the termination prover is abstract in our certification framework.
This means that we are not bound to a specific termination tool.
So we can verify
any termination tool that implements the interpretation method described here
and can output proof traces in \onijn{} format.

Since \wanda{}~\cite{kop:20} is a termination tool that implements the
interpretation method in~\cite{fuhs:kop:12},
it is our first candidate for verification.
We added to \wanda{} the runtime argument
\lstinline[language=bash, style=custombash]{(--formal)}
so it can output proof traces in \onijn{} format.
In~\cite{kop:20} one can find details on how to invoke \wanda{}.
For instance,
we illustrate below how to
run \wanda{} on the \( \map \) AFS\@.
\begin{lstlisting}[language=bash, style=custombash]
  (./wanda.exe) (-d) [rem] (--formal) <Mixed_HO_10_map.afs>
\end{lstlisting}
The setting \lstinline[language=bash, style=custombash]{(-d) [rem]}
sets \wanda{} to disable rule removal.
The option
\lstinline[language=bash, style=custombash]{(--formal)}
sets \wanda{} to only use polynomial interpretations
and output proofs to \onijn{} proof traces.
Running \wanda{} with these options gives us the proof trace we used for
\( \rules_\map \) above.
The latest version of \wanda{}, which includes this parameter,
is found at~\cite{kop:23}.

The table below describes our experimental evaluation on verifying \wanda{}'s
output with the settings above.
The benchmark set consists of those 46 TRSs that \wanda{} outputs \texttt{YES}
while using only polynomial interpretations and no rule removal.
The time limit for certification of each system is set to 60 seconds.

The experiment was run in a machine with
M1 Pro 2021 processor with 16GB of RAM\@.
Memory usage of \nijn{} during certification ranges from 400MB to 750MB\@.
We provide the experimental benchmarks
at~\href{https://github.com/deividrvale/nijn-coq-script-generation}{https://github.com/deividrvale/nijn-coq-script-generation}.

\begin{table}[htb]
  \begin{tabular}{ l | c c c | c c c  }
  & \wanda{} &  & & \nijn{}/\onijn{} & & \\
  \hline
  Technique & \# YES & Pct. & Avg. Time & \# Certified & Perc. &  Avg. Time \\
  \hline
  Poly, no rule removal & 46 & 23\% & 0.07s & 46 & 100\% & 4.06s \\
  \end{tabular}
\caption{Experimental Results}%
\label{table:example}
\end{table}
Hence,
we can certify all TRSs proven SN by \wanda{} using only polynomial interpretations.

\section{Conclusions and Future Work}\label{sec:conclusion}
We presented a formalization of the polynomial method in higher-order rewriting.
This not only included the basic notions, such as algebraic functional systems,
but also the interpretation method
and the instantiation of this method to polynomials.
In addition,
we showed how to generate Coq scripts from the output of termination provers.
This allowed us to certify their output
and construct a formal proof of strong normalization.
We also applied our tools to a concrete instance,
namely to check the output of Wanda.

There are numerous ways to extend this work.
First,
one could formalize more techniques from higher-order rewriting,
such as tuple~interpretations~\cite{kop:vale:21}
and
dependency~pairs~\cite{kop:raamsdonk:12,kus:iso:sak:bla:09}.
One could also integrate HORPO into our framework~\cite{koprowski:09}.
Second,
in the current formalization, the interpretation of application is fixed for every instance of the polynomial method.
One could also provide the user with the option to select their own interpretation.
Third,
currently, only Wanda is integrated with our work.
This could be extended so that there is direct integration for other tools as well.

\bibliography{literature}

\end{document}